\documentclass[aps,prb,superscriptaddress,twocolumn,longbibliography]{revtex4-2}
\usepackage[hidelinks]{hyperref}
\usepackage[T1]{fontenc}
\usepackage{inputenc}
\usepackage{lmodern}
\expandafter\let\csname equation*\endcsname\relax
\expandafter\let\csname endequation*\endcsname\relax
\usepackage{amsmath}
\usepackage{esint}
\usepackage{amssymb}
\usepackage{graphicx}
\usepackage[dvipsnames]{xcolor}
\usepackage{natbib}
\usepackage{bm}
\usepackage{bbold}
\usepackage[mathscr]{euscript}
\usepackage[cal=boondoxo]{mathalfa}
\usepackage{comment}
\usepackage{makecell}
\usepackage{bbding}
\usepackage{pifont}
\usepackage{tikz}
\usetikzlibrary{arrows.meta,
                calc,
                positioning,
                quotes} 
\usepackage[caption=false]{subfig}

\def\pM{\mathrel{\raise 2pt \hbox{\tiny(}\!\raise 1pt \hbox{+}\settowidth {\dimen03} {+}\hskip-\dimen03 \raise -2.4pt \hbox {$-$} \!\raise 2pt \hbox{\tiny)}}}

\begin{document}

\title{Lattice Relaxation Flattens Chern Bands in Rhombohedral Graphene Stacks}

\author{Luca M. Nashabeh}
\email{lmn2163@columbia.edu}
\author{H\'ector Ochoa}
\email{ho2273@columbia.edu} 

\affiliation{Department of Physics, Columbia University, New York, New York 10027, USA}

\begin{abstract}
  Motivated by recent observations of integer and fractional Chern insulators in rhombohedral graphene stacks aligned with hexagonal boron nitride (hBN), we propose and study a model in which the moir\'e potential  is defined by the pattern of layer-shear strain fields produced by lattice relaxation in these heterostructures. Although these strain fields decrease exponentially with the number of layers, their imprints on electrons residing away from the contact layer are non-negligible. In the absence of a displacement field, lattice relaxation effects amplify the electronic differences among the two different stackings with hBN. These differences, although attenuated at the single-electron level, survive in the so-called moir\'e-distant regime and are further enhanced with the inclusion of electron interactions. We find that lattice relaxation plays a crucial role in flattening and isolating a valley-polarized Hartree-Fock electron band with $|C|=1$ Chern number. Our results challenge the conventional wisdom on moir\'e effects in these heterostructures by illustrating the intertwined effects of long-range Coulomb interactions and lattice relaxation, and opens the door to explore different regimes of twist angles and displacement fields for the search for topological states.
\end{abstract}

\maketitle

\section{Introduction}

Fractional Chern insulators (FCIs) \cite{PhysRevX.1.021014,PhysRevLett.106.236804,Sheng:2011aa,PhysRevLett.106.236803,PhysRevLett.106.236802} are crystalline analogs of the fractional quantum Hall effect arising from a spontaneous breaking of time-reversal symmetry rather than the application of a strong magnetic field. Their recent observation in rhombohedral graphene stacks aligned with hexagonal boron nitride (hBN) \cite{Lu:2024aa,Lu:2025aa,Xie:2025aa,Aronson:2025aa,Choi:2025aa,huo2025,xie2025} adds to the discovery of this phenomenon in twisted MoTe$_2$ bilayers \cite{Park:2023aa,Zeng:2023aa}. However, while in twisted MoTe$_2$ the evolution of the moir\'e band topology with twist angle is well understood, in the case of  rhombohedral graphene/hBN superlattices the role of the moir\'e potential is currently under debate \cite{PhysRevLett.133.206502,PhysRevLett.133.206504,guo_fractional_2024, PhysRevB.109.205122,kwan_moire_2025,PhysRevB.110.115146,PhysRevB.112.075110,PhysRevB.111.075130,zlqg-sj86,PhysRevLett.133.206503,PhysRevB.110.205124,PhysRevB.110.205130}.

The observation of FCIs in these heterostructures is striking for several reasons. As in the case of twisted MoTe$_2$, FCIs are observed in concomitance with spin-valley
polarized Chern insulators at integer fillings \cite{Lu:2024aa,Lu:2025aa,Xie:2025aa,Aronson:2025aa,Chen:2020aa,PhysRevX.15.011045,xiang2025continuouslytunableanomaloushall,wang2025electricalswitchingcherninsulators}. However, according to the models in the literature, the Chern insulators do not arise from a parent single-particle topological band, 
pointing to a more prominent role of interactions beyond lifting the spin and valley degeneracies. Moreover, at low temperatures, the Chern insulator at filling $\nu=1$ expands into an \textit{extended} quantum anomalous Hall state \cite{Lu:2025aa,PhysRevX.15.011045,xiang2025continuouslytunableanomaloushall} that in some cases replaces the FCIs. This state has been theoretically identified as a quantum anomalous Hall crystal \cite{PhysRevLett.57.922,PhysRevB.39.8525} in which translational symmetry is spontaneously broken \cite{PhysRevLett.133.206503,PhysRevB.110.205124,PhysRevB.110.205130}, relegating the moir\'e potential to a secondary role. Despite recent systematic investigations of electron crystal phases in this system \cite{xq6w-96vg}, the extended impact of layer-shear strain fields on these Hartree-Fock bands has remained unexplored until now. Also from the theory side, it has been noted that the FCIs are numerically unstable \cite{PhysRevB.112.075110,zlqg-sj86} owing to the unavoidable multi-band character of the problem. This numerical observation also points to the shortcomings of calculation schemes and/or current models in the literature.

These sometimes conflicting views and observations have motivated us to revisit the single-particle model of rhombohedral graphene/hBN superlattices. On phenomenological grounds, Dirac electrons from two inequivalent valleys $K_{\pm}$ experience scalar, staggered, and pseudo-gauge potentials modulated on the moir\'e scale defined by the lattice constant mismatch $\delta$ and twist angle $\theta$ \cite{wallbank_generic_2013,mucha-kruczynski_heterostructures_2013}. In most models, these potentials arise from second-order processes in the electronic tunneling between graphene and hBN \cite{jung_ab_2014,kindermann_zero-energy_2012,moon_electronic_2014}. 
However, it was pointed out more than a decade ago 
that close to perfect alignment, strong lattice distortions develop in graphene due to the moir\'e adhesion landscape, modifying the electronic properties; for example, a spectral gap is developed in single-layer graphene due to unequal expansion of partially commensurate regions of the moir\'e pattern \cite{san-jose_electronic_2014,san-jose_spontaneous_2014}.

In this work we examine the role of the moir\'e potential created by these lateral tensions. Unlike other approaches \cite{PhysRevB.109.205122}, in our model lattice relaxation is not introduced as a reparametrization of the tunneling parameters between hBN and the contact graphene layer, but as a lattice mismatch that propagates over several layers, producing changes in the interlayer tunneling as well as in the intralayer hopping terms. The latter gives rise to pseudo-magnetic fields $\mathscr{B}$ of the order of T in the contact layer for $\theta=0.77^{\circ}$, which decay quickly in the number of layers, although they still produce an appreciable effect, e.g., on the second layer, where $\mathscr{B}$ is of the order of hundreds of mT.

With our model we find that lattice relaxation enhances the electronic differences between the two possible stacking configurations with hBN --despite the fact that the strain fields are the same for both configurations. These differences persist in the so-called \textit{moir\'e distant} regime, when the displacement field pushes conduction electrons away from the contact layer, and are further enhanced by long-range Coulomb interactions. 

Both lattice relaxation and Coulomb interactions are crucial for flattening and isolating the Chern band that hosts the FCIs. This conclusion is supported by self-consistent Hartree-Fock calculations for the valley-polarized state that results from adding one electron per moir\'e cell in a (spinless) interacting model where relaxation effects can be switched off. Our results suggest that either of the two possible stackings close to the experimentally reported twist angle of $0.77^\circ$ are optimal for hosting FCIs, but do not exclude the possibility of other combinations of stacking, twist angle, and displacement field.

\begin{figure}
    \centering
    \scalebox{0.59}{\begin{tikzpicture}
    \newdimen \R \R=2cm
    \begin{scope}[shift = {(0\R, -1\R)}]
        \def\d{0.85*\R}
        \foreach \x/\y in {0/1,1/2,2/3,0/4,1/5}
        {\draw[gray] (0, {(\y-1) * \d/2}) -- (3*\d, {(\y-1) * \d/2});
        \draw[ultra thick] ({\x * \d},{(\y-1) *\d/2} ) -- ({(\x+1) * \d}, {(\y-1) *\d/2}) node[fill = black, circle, pos = 0] {} node[fill = black, circle, pos = 1] {} ;
        \node at (-0.4*\d, {(\y-1)*\d/2}) {\Large $\bm u_{\y}$};
        \node at (1.5*\d, {(\y - 1.5) * \d/2}) {\Large $\bm \phi_{\y}$};
        }

        \node[above] at (1 * \d, 2.1*\d) {\LARGE $C_A$};
        \node[above] at (2 * \d, 2.1*\d) {\LARGE $C_B$};
        
        \draw[gray] (0, -\d/2) -- (3*\d, -\d/2);
        \draw[ultra thick] (0,-\d/2 ) -- (\d, -\d/2) node[fill = blue, circle, pos = 0] {} node[fill = red, circle, pos = 1] {};

        \node[below, blue] at (0, -0.6*\d) {\LARGE $B$};
        \node[below, red] at (\d, -0.6*\d) {\LARGE $N$};

        \node[] at (2.4*\d, -0.8*\d) {\LARGE $\eta = +1$};
    \end{scope}

    \draw[thick, dashed] (2.95\R, -2\R) -- (2.95\R, 1.4\R);

    \begin{scope}[shift = {(5\R, 0)}]
        \def\d{1.2}
        \def\t{15}

        \foreach \x in {0,60,...,300}
        \draw[thin] (\x:{sqrt(3)*\R/\d/2}) -- (\x+60:{sqrt(3)*\R/\d/2});

        \foreach \x in {0,60,...,300}
        \draw[line width=1.5pt] (\x+\t:{sqrt(3)*\R/2}) -- (\x+60+\t:{sqrt(3)*\R/2});
        
        \foreach \x/\l/\c in {30/1/blue, 150/2/olive, 270/3/red}
        {\draw[line width=2.5pt,\c, opacity=0.6] (0,0) edge[-stealth,"\large $\bm h_\l$"', pos=0.9] (\x:{sqrt(3)*\R/\d});
        \draw[line width=2.5pt,\c] (0,0) edge[-stealth,"\large $\bm  g_\l$", pos=0.9] (\x+\t:{sqrt(3)*\R});
        \draw[line width=1.5pt, -stealth,\c] (\x+\t:{sqrt(3)*\R}) -- (\x:{sqrt(3)*\R/\d});
        \node[\c] at (\x+\t/3:{1.05*sqrt(3)*\R}){\large $\bm G_\l$};
        }

        \draw[very thick, -stealth] (270:0.9*\d*\R) arc (270:270+\t:\d*\R) node[midway, below] {\Large $\theta$};

        \filldraw[black] (\t:{sqrt(3)*\R/2}) circle (\R/15) node[below right] {\large $\bm K_+$};
        \filldraw[black] (180+\t:{sqrt(3)*\R/2}) circle (\R/15) node[below left] {\large $\bm K_-$};
    \end{scope}
\end{tikzpicture}}
    \caption{The geometry of the system consists of five layers of rhombohedral graphene stacked on an hBN layer. In the $\eta = +1$ configuration, $C_A$ is above boron, while in the $\eta = -1$ $C_A$ is above nitrogen. In momentum space, the moir\'e reciprocal vectors arises from displacements between the graphene and hBN reciprocal vectors (figure is not to scale). The two $K_{\pm}$ valleys are indicated. }
    \label{fig:geometry}
\end{figure}
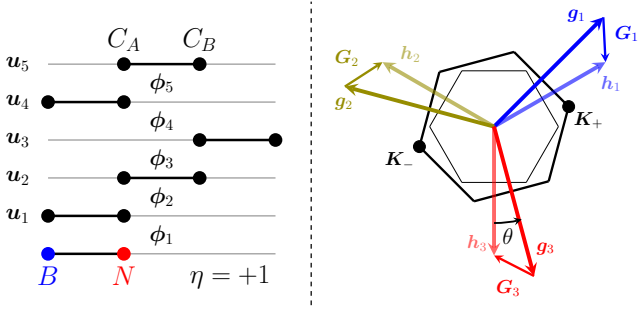

\begin{figure*}[!t]
    \centering
    \includegraphics[width=0.99\linewidth]{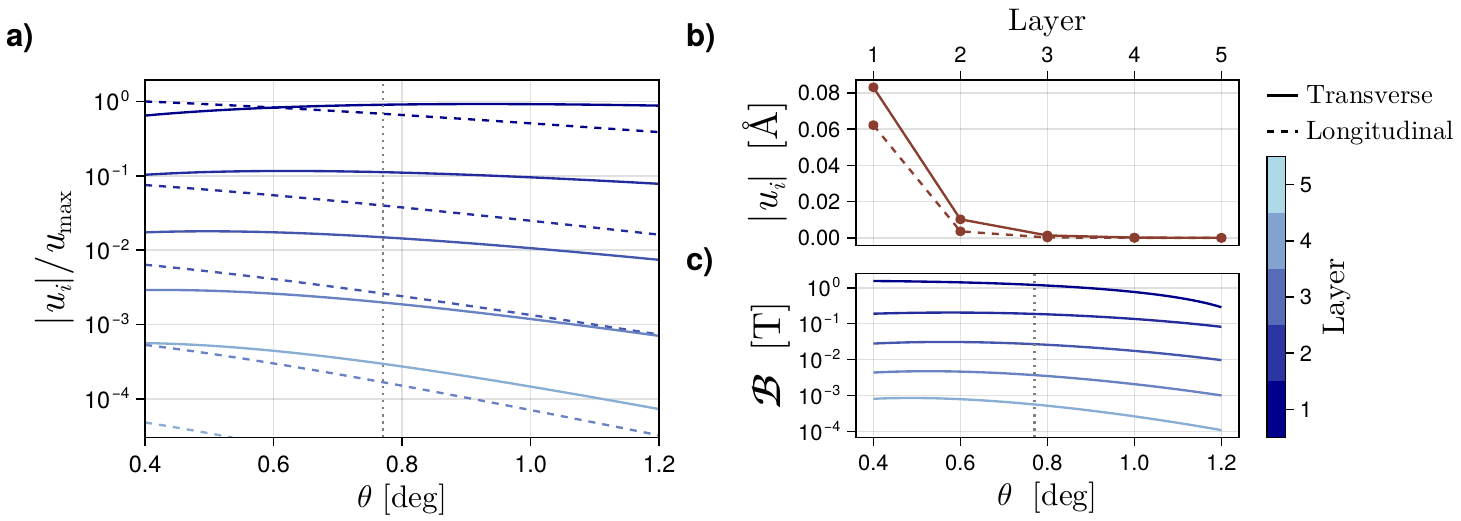}
    \caption{Results of relaxation for $\eta = 1$. \textbf{a)} The transverse (continuum lines) and longitudinal (dashed lines) components of the relaxation displacement fields have different behaviors as functions of angle. While the longitudinal field monotonically decreases with twist angle, the transverse field has maxima at different critical angles depending on the layer number $n$. \textbf{b)} As a function of layer number, both fields show an expected exponential falloff, with characteristic length $l\sim 1.6$ {\AA} at $\theta = 0.77^\circ$. \textbf{c)} The pseudomagnetic field $\mathscr{B}$ is concentrated in the contact layer, and depends strongly on angle only in the farthest layers. $\eta = -1$ is qualitatively similar, with negligible differences in the magnitude of relaxation effects.}
    \label{fig:relaxation}
\end{figure*}

\section{Lattice relaxation}

In rhombohedral systems the graphene layers are stacked in a metastable staircase configuration as schematically represented in the left panel of Fig.~\ref{fig:geometry}. The partial alignment with a hBN substrate reduces the symmetry to a vertical 3-fold rotation axis. The lattice mismatch and twist angle produces a moir\'e beating pattern with vectors $\boldsymbol{G}_j$ defined in the right panel of Fig.~\ref{fig:geometry}. For the same set of vectors $\boldsymbol{G}_j$, there are two possible stacking configurations with hBN (hereafter labeled by index $\eta=\pm 1$) related by a relative 180$^{\circ}$ rotation with respect to the graphene layers.

Lattice relaxation results from the competition between elastic energy within the graphene layers---which resists excessive displacements---and the interlayer adhesion potential---which drives the twisted layers to realign. This leads to a local displacement $\bm u_l$ in the $l$th layer, counting from the closest to hBN, see Fig.~\ref{fig:geometry}. 
It is also convenient to introduce the local stacking configuration of layer $l$ as
\begin{equation}
    {\bm \phi}_l(\bm r) = \begin{cases}
    (1-\delta\, R_\theta) \bm r + \bm u_1(\bm r) & \textrm{for}\,\,l=1,\\
    \bm u_l(\bm r) - \bm u_{l-1}(\bm r) &\textrm{for}\,\, l>1,
\end{cases}
\end{equation}
which describes the relative displacements between adjacent layers, including the macroscopic twist displacement between hBN and the first graphene layer. $\delta = a_{\text{G}}/a_{\text{hBN}}$ determines the lattice mismatch, while $R_\theta$ is the relative rotation. In terms of these, the interlayer adhesion potentials can then be expressed as
\begin{equation}
    \begin{split}
        \mathscr V_1 &= 2\mathrm{Re}\left[V_{\text{hBN}}\sum_{j=1}^3 e^{i \bm h_j \cdot \bm \phi_1}\right],\\
        \mathscr V_{l} &= 2\mathrm{Re}\left[V_{\text{G}}\sum_{j=1}^3 e^{i \bm g_j \cdot \bm \phi_l}\right] \qquad\quad \,\,\, \textrm{for}\,\,l>1.\\
    \end{split}
\end{equation}

However, when solving for the final relaxed configuration, we find  more convenient to work directly with the displacements $\bm u_l$. The lattice relaxed configuration is then set by the condition 
\begin{equation}\label{eq:EL-relax}
    \frac{\lambda + \mu}{2}\nabla(\nabla \cdot {\bm u}_l) + \frac{\mu}{2}\nabla^2 {\bm u}_l = \frac{\delta \mathscr V_{\text{total}}}{\delta{\bm u}_l},
\end{equation}
where $\mathscr V_{\text{total}} $ is the sum of all adhesion potentials, $\mathscr{V}_n$, and $\lambda,\mu$ are the Lam\'e coefficients of graphene. The various parameters used for this model are given below in Table~\ref{tab:relaxation}.

Since the resulting displacements $\bm u_l(\bm r)$ are periodic with respect to the moir\'e lattice, we can further decompose them in longitudinal and transverse harmonics,
\begin{equation}
    \bm u_l(\bm r) = \sum_{\bm G_j} (u_{j, l}^L \bm{\hat G}_j + u_{j, l}^T \bm{\hat G}_j \times \bm{\hat z}) e^{i\bm G_j \cdot \bm r}.
\end{equation}
This decomposition is particularly convenient for solving Eq.~\eqref{eq:EL-relax} self-consistently. For our calculations, we only include the first six harmonics with $|\bm G_j| = \bm G_1$. Because of the three-fold rotational symmetry of the system and the requirement that displacements be real, any one of these six harmonics fully determines the rest. We numerically find higher harmonics to have little effect on the resulting band structure. 

\begin{table}[!b]
    \centering
    \def\arraystretch{1.1}
    \setlength\tabcolsep{1ex}
    \begin{tabular}{c|c l|l}
         Variable & Value & & Source  \\\hline
         $a_{\text{G}}$ & 2.46 & \AA & \cite{san-jose_spontaneous_2014} \\
         $a_{\text{hBN}}$ & 2.504 & \AA & \cite{san-jose_spontaneous_2014} \\
         $d$ & 3.33 & \AA & \cite{kwan_moire_2025}\\
         $\lambda$ & 3.25 & eV/\AA$^2$ & \cite{san-jose_spontaneous_2014}\\
         $\mu$ & 7.8 & eV/\AA$^2$ & \cite{san-jose_spontaneous_2014} \\\hline
         $V_\text{G}$ & $4.444\,e^{2\pi i/3}$ & meV/\AA$^2$  & \cite{ochoa_degradation_2022}\\
         $V_{\text{hBN}}$ & $1.166 - \eta 1.652i$ & meV/\AA$^2$ & \cite{san-jose_spontaneous_2014}
    \end{tabular}
    \caption{Parameters used for the lattice relaxation model.}
    \label{tab:relaxation}
\end{table}

The results of lattice relaxation in pentalayer rhombohedral graphene across a range of twist angles are given in Fig.~\ref{fig:relaxation}. As expected, the displacement field $\bm u_n$ is largest in the first layer, as this is the only layer which directly experiences the hBN adhesion potential driving the relaxation. Subsequent layers have an exponentially suppressed tendency to match this displacement, as rhombohedral stacking is a locally stable stacking configuration.  

We see two different behaviors for the longitudinal and transverse displacements. The longitudinal displacement monotonically decreases with twist angle, while the transverse displacement field achieves a maximum at a non-zero layer-dependent maximum angle. This can be understood since, at $\theta = 0$, the longitudinal component solely compensates for the lattice mismatch between graphene and hBN, while the transverse field vanishes. As twist angle increases, lattice mismatch becomes less relevant, while the transverse field begins compensating for the twist mismatch. Numerically, we find the transverse field maximum to be around $\theta \approx 0.9^\circ $ in the first layer and $\theta \approx 0.6^\circ$ in the second layer, both near the experimental angle of $0.77^\circ$. 

\section{Single-Electron Model}
\begin{figure*}[t]
    \centering
    \includegraphics[width=0.99\linewidth]{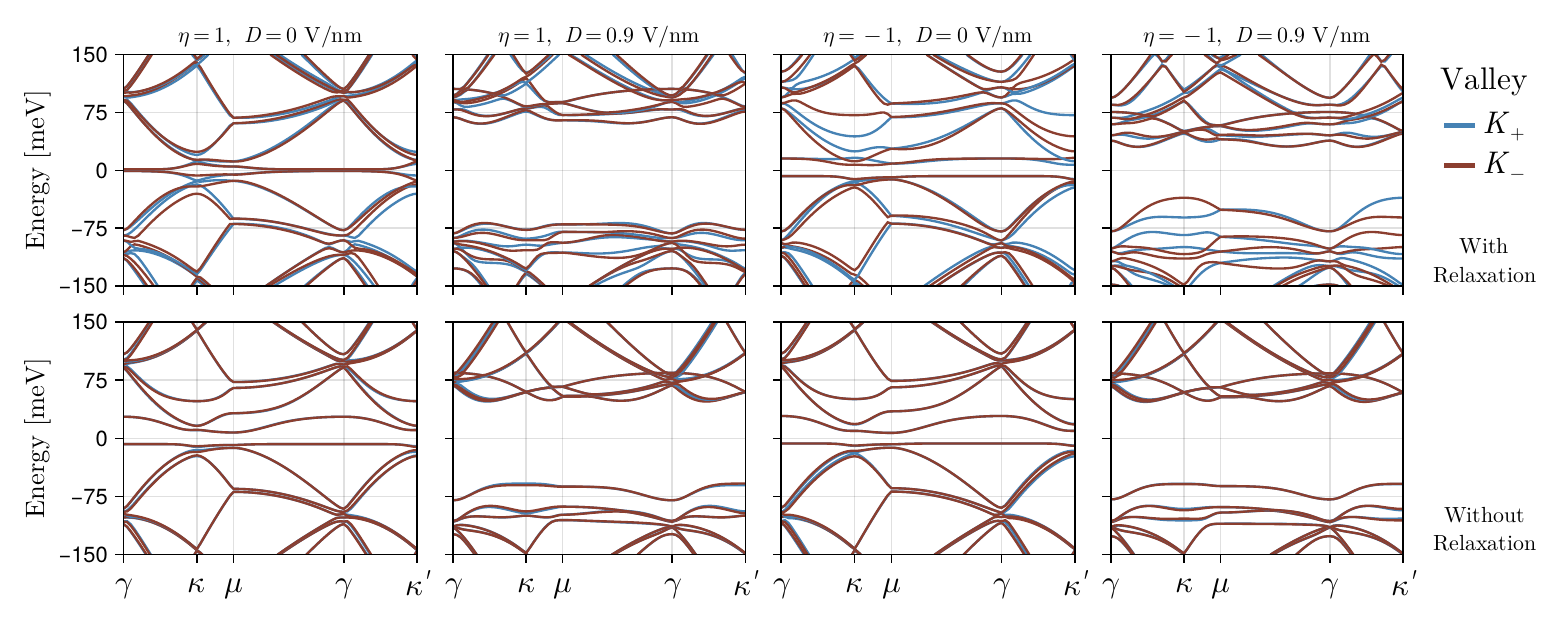}
    \caption{Calculated single-electron band structure for different stacking configurations and displacement fields, plotted over the moir\'e Brillouin zone. The top row includes strain relaxation, while the bottom row does not. Note that strain relaxation introduces appreciable differences between the two valleys.}
    \label{fig:single-particle}
\end{figure*}

The effects of lattice relaxation manifest as effective modifications to both the intra and interlayer hoppings. For the intralayer hoppings, the primary impact of lattice relaxation is captured by the introduction of a pseudogauge field in the intralayer $K_\pm$-valley Hamiltonian,
\begin{equation}
    H_{l}^{\pm} = - v_f (-i\hbar\nabla-e\bm{\mathscr A}_l^{\pm}) \cdot (\pm \sigma_x, \sigma_y) + \frac{eD}{\epsilon}ld,
\end{equation}
where $\sigma_i$ are Pauli matrices acting on sublattice and we are folding momenta to the moir\'e Brillouin zone, with the original graphene Dirac points lying at the zone center. Explicitly, $\bm{\mathscr A}_n^{\pm}$ reads
\begin{equation}
     \bm{\mathscr A}_{l}^{\pm} = \frac{\pm t\beta}{e v_F}\begin{pmatrix}
         u_{l\,xx}-u_{l\,yy}\\-2u_{l\,xy},
     \end{pmatrix}
\end{equation}
where $u_{l\,ij} = (\partial_i u_{l\,j} + \partial_j u_{l\,i})/2$ are the strain tensor components in layer $n$ and $\beta = {\mathrm{d} \log t}/{\mathrm{d} \log a_{\text{G}}}\approx 2$ measures the effect of lattice distortion on hopping \cite{san-jose_spontaneous_2014}. The resulting pseudomagnetic field strength $\mathscr{B}$ is also plotted in Fig.~\ref{fig:relaxation}. Notably, the field in the first layer is of order $\mathscr{B_1} \sim 1\;\text{T}$. 
A scalar divergence term $V_l = V_d\,  u_{l\, ii}$ should also be considered in the effective description of lattice relaxation, however screening is known to effectively set $V_d = 0 $ \cite{park_electronphonon_2014, sohier_phonon-limited_2014}. For the interlayer hoppings, we follow the methods of Moon and Koshino \cite{moon_electronic_2014} and include the effects of the moir\'e and relaxation as additional phase factors in the interlayer hopping caused by the local horizontal displacements. We also absorb hoppings through the hBN layer as an effective intralayer term in the first graphene layer. The resulting Hamiltonian is described in detail in the Supplementary Information. 

We plot the single-particle band structure at $D=0$ and $D=0.9\;\text{V}/\text{nm}$ for the two different stacking configurations $\eta = \pm 1$ both with and without lattice relaxation in Fig.~\ref{fig:single-particle}. Our semi-analytical relaxation results in a qualitatively similar band structure to those produced with previous atomistic dynamic models \cite{guo_fractional_2024, kwan_moire_2025}. Lattice relaxation already has a large impact at the single-particle level, introducing significant modifications to the conduction and valence bands that depend quite dramatically on $\eta$. These modifications suggest that lattice relaxation plays an important role in the detailed physics of pentalayer rhombohedral graphene. 

For $\eta =1$ and $D=0$, the primary impact is to significantly decreases the band gap---flattening the first conduction band and bringing it to almost touch the valence band---and creating band crossings between the first and second conduction/valence bands. At larger displacement fields, the effect of lattice relaxation is instead to squeeze the valence side bands together, while opening a very slight gap for the first conduction band. This is interesting, as it is ultimately the first conduction band that is known to experimentally host fractional Chern states. Lattice relaxation also introduces a slight splitting between the two valleys $K_{\pm}$. 

In contrast, for $\eta = -1$ the band structure is not as dramatically shifted after the introduction of lattice relaxation. For $D=0$, there is a slight narrowing of the first conduction band, while at large displacement field the first valence band remains gapped from the further valence bands. Instead, the $\eta=-1$ lattice relaxed band structures show a significantly greater degree of valley splitting, with large differences between certain $K_\pm$ bands near the $\kappa$ and $\kappa'$ points in the moir\'e Brillouin zone, especially on the hole side.

\begin{figure}[t]
    \centering
    \includegraphics[width=0.99\linewidth]{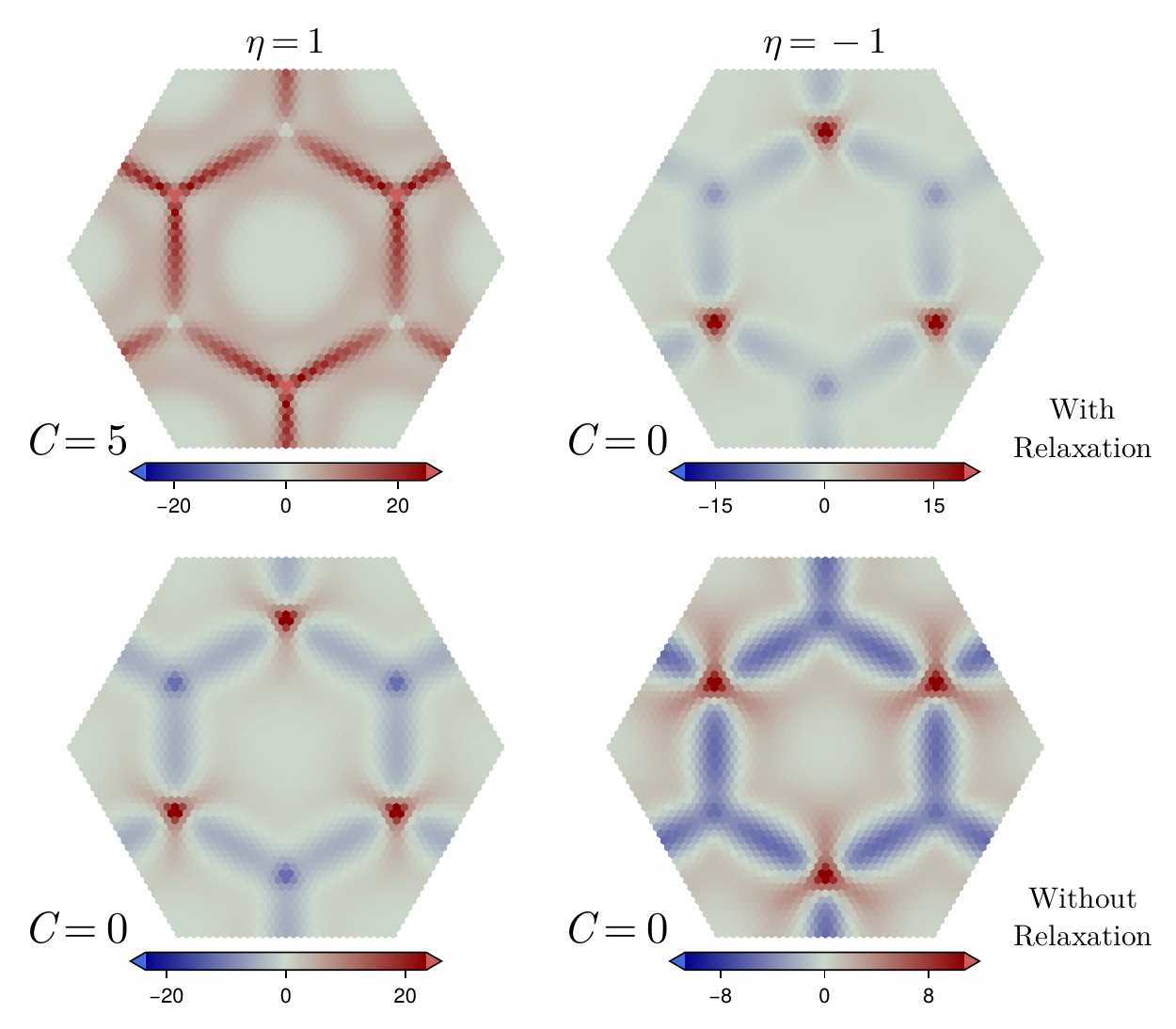}
    \caption{$K_{+}$ first conduction band Berry curvature $\Omega(\bf k)$ in the relaxed (upper) and unrelaxed (lower) cases for $D=0$. In particular, in the $\eta=1$ stacking configuration, relaxation sufficiently perturbs the band structure, resulting in a non-zero Chern number at the single-particle level. Three reciprocal unit cells are shown, and the color scale is given by $\Omega({\bf k}) \cdot A_{\text{MBZ}}/(2\pi)$. }
    \label{fig:sp-chern}
\end{figure}

We also calculate the Berry curvature and Chern number of the conduction band for some paradigmatic scenarios, with the results shown in Fig.~\ref{fig:sp-chern}. We again see that lattice relaxation introduces significant differences between the two stacking configurations, while in the unrelaxed model the differences are largely captured by an inversion. Most strikingly, while the Chern number generally vanishes at the single-particle level, we find that for $\eta = 1$ lattice relaxation alone is enough to induce a non-zero Chern number at zero displacement field. This not only shows that the effects of lattice relaxation cannot be simply described as an adiabatic modification to the bands, but further suggests that they play a key role in the Hall physics. The addition of weak displacement fields does not significantly alter these results, although at strong fields the conduction band is generally no longer isolated in the unrelaxed model and its Chern number cannot be defined. 

\section{Hartree-Fock Calculations}
\begin{figure*}[!t]
    \centering
    \includegraphics[width=0.99\linewidth]{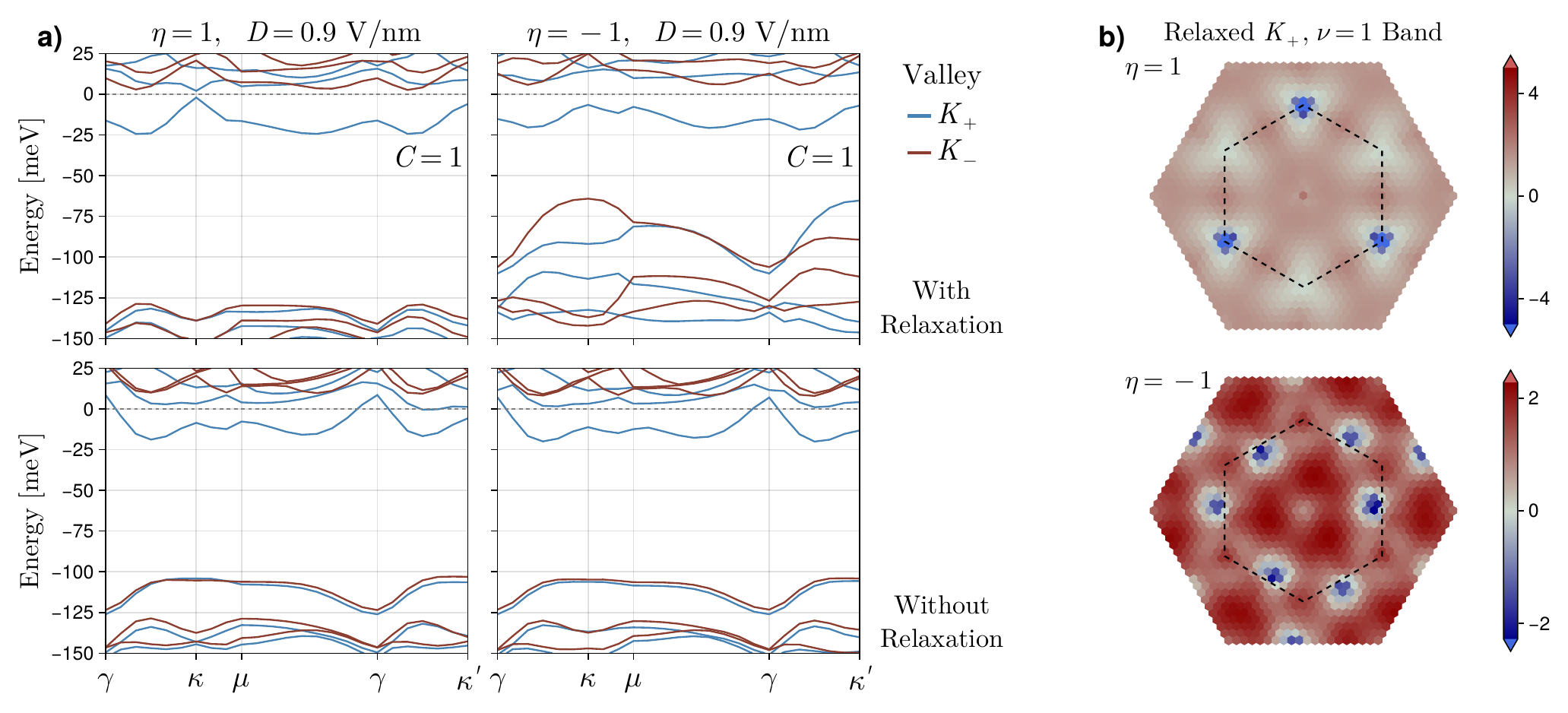}
    \caption{\textbf{a)} The Hartree-Fock single-electron spectrum with and without relaxation effects for filling $\nu=1$ (spinless). Fully including electronic interactions in the moir\'e distant regime, relaxation is essential for the development of a flat valley-polarized electron band. The difference between the two stacking configurations is only significant with relaxation effects. An $18\times 18$ $k$-mesh is used, with $\pm 3$ active bands around the Fermi level per valley. \textbf{b)} The Berry curvature of the relaxed Hartree-Fock filled conduction band; three unit cells are shown with the 1st Brillouin zone marked. The further flattening of the $\eta = -1$ stacking is manifest in its smoother curvature. A $20\times 20$ $k$-mesh is used with all other parameters the same. 
    }
    \label{fig:all_HF}
\end{figure*}

The combination of long-range Coulomb interactions and lattice relaxation amplify the differences between the two different stackings even in the moir\'e-distant regime. In order to study their interplay, we include an interaction term in our Hamiltonian consisting of layer-resolved matrix elements of a dual-gate screened Coulomb potential \cite{kwan_moire_2025}. After a standard Hartree-Fock decoupling of this term and neglecting the electron spin, we look for valley-polarized self-consistent solutions when one electron per moir\'e cell is added to the system (details of the calculation can be found in the Supplementary Information). We refer to this situation as filling $\nu=1$ in our spinless model. Importantly, in our calculations we compare the model with and without relaxation effects, for the latter making $\boldsymbol{u}_l=0$ everywhere. 

The main conclusion is that relaxation contributes to isolate and flatten a topological, valley-polarized electron band. A representative set of results is displayed in Fig.~\ref{fig:all_HF}, where the Hartree-Fock bands in the moir\'e distant regime ($D=0.9$ V/nm) are displayed for both stackings (left and right in panel~a). The comparison of the results with and without relaxation (top and bottom figures, respectively) shows that relaxation is crucial to avoid a band touching at the zone center. In the case of stacking $\eta=-1$, relaxation effects give rise to a very flat electron band well separated from the rest of conduction bands, giving rise to a smoother distribution of the Berry curvature in panel~b. The Chern number is $C=1$ (see details of the calculation in the Supplementary Information, also Ref.~\cite{PhysRevB.86.115112}). In the case of stacking $\eta=+1$, the result is qualitatively the same, but we find a pronounced dispersion of the topological electron band around the zone corners, where the Berry curvature is concentrated.

In concomitance with lattice relaxation, long-range Coulomb interactions contributes to flatten and define the topological character of the electron band --in addition to promoting the symmetry breaking. We studied how the Hartree-Fock bands of our model evolve as a function of the dielectric environment (parametrized by the relative permittivity $\epsilon$). We saw (see the Supplementary Information) that in the relaxed $\eta=+1$ stacking the value of $\epsilon$ drives a topological transition through a band touching of the filled electron band with the empty conduction band at the zone corners. The transition takes place within the range $\epsilon=5.5-6$, and the Chern number of the electron band changes from $C=1$ to $C=2$. This is not the case for the other stacking, $\eta=-1$, at least within the parameter range that we explored. However, in both cases the suppression of Coulomb interactions with $\epsilon$ produce more dispersive electron bands, as well as less isolated from the rest of the single-particle spectrum. Furthermore, decreasing Coulomb repulsion tends to restore the resemblance in the spectrum of both stackings in the moir\'e-distant regime. This exemplifies how lattice relaxation and long-range Coulomb interactions are intertwined.

\section{Discussion and conclusions}
In this article, we have introduced a method for quantitatively including the effects of relaxation directly into the Hamiltonian for moir\'e rhombohedral stacks, accounting for both the standard effects of virtual interlayer hopping through the hBN substrate and the effects of real changes in the intralayer hopping from moir\'e elastic distortions. Not only do we find the magnitude of the relaxation displacements to be significant over a range of angles $\theta \lesssim 1^\circ$, but the corresponding pseudomagntic field has magnitude of order $\mathcal B \sim 1 \text{ T}$ in the lowest layers. These effects begin decay quickly for larger twist angles, matching the experimental necessity of small twist angles. 

At the single-particle level, relaxation effects already play a key role in determining the low energy physics, causing significant rearrangement of the bands that go beyond an adiabatic shift. This is especially true at zero displacement field, though even at large displacement fields relaxation already helps isolate the first conduction band and introduces local valley splitting.

Although the relaxation displacements are agnostic of the specific stacking $\eta=\pm 1$, their imprint on the electronics amplify their differences even in the moir\'e-distant regime at finite fillings if long-range Coulomb interactions are taken into account. 
This is due to the subtle interplay between relaxation and electrostatic effects through the spatial distribution of the excess charge and the associated Hartree potential.
In fact, in self-consistent Hartree calculations (when we drop the Fock exchange potential, see the Supplementary Information) at $\nu=1$, relaxation always contributes to isolate the partially-filled conduction band (more so for stacking $\eta=+1$). Furthermore, in the moir\'e-proximal regime (with negative $D$ fields in our geometry), the Hartree term is enough to produced valley bands with $|C|=1$ (see the Supplementary Information).

Our results bring into question the conventional wisdom that rhombohedral graphene stacks almost aligned with hBN are largely independent of the detailed effects of the moir\'e pattern through relaxation.
It is true that lattice distortions decay exponentially with layer number, and that effectively the moir\'e hopping term only directly enters in the contact layer Hamiltonian. However, when long-range Coulomb interactions are included, the occupied valence states --which live near the contact layer in the moiré-distant regime and are strongly influenced by both these effects-- can still induce significant changes in the electron bands in the experimentally relevant regime of displacement fields. Because these relaxation effects contribute directly to isolating the topological electron band, faithfully capturing them provides a more reliable starting point for understanding the numerical stability of FCI states.

In closing, we want to emphasize again the relevance of having an accurate model for the strength of the moir\'e potential. Indeed, the details of the transition between a strong moir\'e and an approximately bulk rhombohedral graphene could be important in explaining the observation of the extended $\nu=1$ quantum Hall phase. Specifically, one could imagine that albeit the moir\'e periodicity determines the densities needed for exact integer fillings, the exponential decay in lattice distortions could allow electrons in the top layers to rearrange and maintain an integer filling at nearby densities with little energy cost.

\begin{acknowledgments}
\textit{Acknowledgments}---L.M.N was supported by the Columbia University I.I. Rabi Scholarship. 
\end{acknowledgments}

\bibliography{biblio_rh5Gr+hBN}

\onecolumngrid
\section*{Supplementary Information}




\subsection{Details of Single-Electron Hamiltonian}
Letting $\xi=\pm 1$ be the valley index with $\bm K_\xi = \xi (\bm g_1 - \bm g_2)/3$ (see definitions in Fig.~\ref{fig:geometry} of the main text), and $\eta=\pm 1$ be the ABC stacking direction index, we may initially write the block Hamiltonian in layer space as
\[H_0 =\begin{bmatrix}
    H_{hBN}  & U          & 0          & 0          & 0          & 0 \\
    U^\dagger& H_{1}      & V_1        & 0          & 0          & 0 \\
    0        & V_1^\dagger& H_{2}      & V_2        & 0          & 0 \\
    0        & 0          & V_2^\dagger& H_{3}      & V_3        & 0 \\
    0        & 0          & 0          & V_3^\dagger& H_{4}      & V_4 \\
    0        & 0          & 0          & 0          & V_4^\dagger& H_{5}
\end{bmatrix}.\]
The diagonal blocks are the intralayer Hamiltonians in sublattice space, and can be expressed, for $\eta = +1$, as 
\[H_{\text{hBN}} = \begin{bmatrix}V_B & 0 \\ 0 & V_N\end{bmatrix}\qquad\text{and}\qquad H_{l} = -v_f (-i\hbar \nabla-e \bm A_l) \cdot (\xi \sigma_x, \sigma_y) + V^d_l(\bm r) + \frac{eD}{\epsilon}nd.\]
$\bm A_l$ and $V^d_l$ are the strain-induced pseudo-gauge field and local deformation potential in each layer, 
\[\bm A_l = \frac{\xi t\beta}{e v_F}\begin{pmatrix}u_{l\,xx}-u_{l\,yy}\\-2u_{l\,xy}\end{pmatrix} \qquad \text{and} \qquad V^d_l = V_d\,{u_{l\,ii}}.\]
\textit{A priori} $V_d$ is non-zero, however, due to screening we take an effective value $V_d =0$ \cite{park_electronphonon_2014, sohier_phonon-limited_2014}. $D$, on the other hand, is any applied out of plane displacement field, with $d$ the graphene interlayer distance. For $\eta = -1$, $V_B$ and $V_N$ are swapped.

The graphene-hBN tunneling block $U$ is generically given in reciprocal space by the following matrix elements in sublattice space:
\[U_{ij}(\bm k, \bm \delta) = \sum_{n,m} \tilde t(n \bm g_1 + m \bm g_2 + \bm k) e^{i(n\bm g_1 + m\bm g_2 + \bm k)(\bm \delta + \bm \tau_{ij})},\]
where $\bm \tau_{ij}$ is the interlayer sublattice separation between site $j$ in the graphene and site $i$ in the hBN, and $\bm \delta$ parametrizes the lattice mismatch due to the moir\'e and additional relaxation effects. In our case, the only relevant tunneling events are from $\bm K_\xi$ to its three equivalent copies, meaning the sum can be reduced to 
\[(n,m) = \{(0,0), (-\xi, 0), (0, \xi)\}.\]
We also note that $\bm \tau_{BA} = -\bm \tau_{AB} = (\bm a_1 + \bm a_2)/3$, where $\bm a_i$ are primitive vectors of graphene's Bravais lattice, while $\bm \tau_{AA} = \bm \tau_{BB}=0$. Substituting $\bm \delta_1 = \bm \delta_{\text{moire}} - \bm u_1$, where $\bm\delta_{\text{moire}}$ describes the moir\'e pattern in the absence of relaxation, and noting that $\bm g_i \cdot \bm \delta_{\text{moire}}=\bm G_i \cdot \bm r$, we conclude that
\[U = u_0 e^{i\bm K_\xi \bm \delta_1}\left(\begin{bmatrix}1 & 1 \\ 1 & 1\end{bmatrix} + \begin{bmatrix}1 & \omega^{\xi}\\\omega^{-\xi} & 1\end{bmatrix} e^{-i\xi (\bm G_1 \bm r - \bm g_1 \bm u_1)} + \begin{bmatrix}1 & \omega^{-\xi}\\\omega^{\xi} & 1\end{bmatrix} e^{i\xi (\bm G_2 \bm r -\bm g_2 \bm u_1)}\right),\]
where $u_0 = \tilde t(\bm K_\xi)\approx {0.152}\;\text{eV}$ and $\omega = e^{2\pi i/3}$.


The matrix elements of the graphene interlayer tunneling terms $V_l$ (which connects layer $l+1$ to layer $l$) are similarly written as
\[V_{l\,ij}(\bm k, \bm \delta) = \sum_{n,m} \tilde t(n \bm g_1 + m \bm g_2 + \bm k) e^{i(n\bm g_1 + m\bm g_2 + \bm k)(\bm \delta + \bm \tau_{ij})},\]
but now we have (in the case $\eta=1$) $\bm \tau_{BA} = 0$, $\bm \tau_{AA}=\bm \tau_{BB} = -\bm \tau_{AB} = -(\bm a_1 + \bm a_2)/3$.
Thus, substituting $\bm \delta_l = \bm u_l - \bm u_{l+1} = \Delta \bm u_l$ gives
\[V_{l} = v_0 e^{i\bm K_\xi \Delta \bm u_l} \left(\begin{bmatrix}1 & 1 \\ 1 & 1\end{bmatrix} + \begin{bmatrix}\omega^\xi & \omega^{-\xi}\\1 & \omega^\xi\end{bmatrix} e^{-i\xi  \bm g_1 \Delta \bm u_l} + \begin{bmatrix}\omega^{-\xi} & \omega^{\xi}\\1 & \omega^{-\xi}\end{bmatrix} e^{i\xi \bm g_2 \Delta \bm u_l}\right).\] 
We note a slight simplification can be made here:
\[e^{i\bm K_\xi \bm \delta_l} e^{-i\xi \bm g_1 \Delta \bm u_l}  = e^{i\xi(\bm K_+ - \bm g_1)\Delta \bm u_l}.\]
Also, it is good to verify that the constant order expansion of the above gives
\[V_{l} \approx  3v_0 \begin{bmatrix}0 & 0 \\ 1 & 0\end{bmatrix},\]
as should be expected. 

Since the hBN layer can be assumed static, we use second-order perturbation theory to absorb its electronic impacts into an effective intralayer term in the 1st layer, giving
\[H_1^{\text{eff}}= H_1 - U^\dagger H_{\text{hBN}}^{-1} U = H_1 + U_{\text{hBN}}.\]
Explicitly, we get
\begin{align*}
    U_{\text{hBN}} &= u'_0\begin{bmatrix}1 & 0 \\ 0 & 1\end{bmatrix}+ u'_1 e^{i\xi \psi} \Bigg(\begin{bmatrix}1 & \omega^\xi \\ 1 & \omega^\xi\end{bmatrix}e^{-i\xi (\bm G_1 \bm r - \bm g_1 \bm u_1)} + \begin{bmatrix}1 & 1 \\ \omega^\xi & \omega^\xi\end{bmatrix}e^{-i\xi (\bm G_2 \bm r - \bm g_2 \bm u_1)} \\
    &\phantom{=============}+ \begin{bmatrix}1 & \omega^{-\xi} \\ \omega^{-\xi}  & \omega^\xi\end{bmatrix}e^{i\xi( (\bm G_1 + \bm G_2) \bm r - (\bm g_1 + \bm g_2) \bm u_1)} \Bigg)+ \text{h.c.}
\end{align*}
where, for $\eta = +1$
\[u'_0 = -3u_0^2 \left(\frac{1}{V_B} + \frac{1}{V_N}\right) \qquad\text{and}\qquad u'_1 e^{i\psi}= -u_0^2 \left(\frac{1}{V_B} + \frac{1}{\omega V_N}\right).\]
For $\eta = -1$, $V_B$ and $V_N$ are swapped.

We thus have an effective continuum Hamiltonian for the active graphene degrees of freedom given by
\[H^{\text{eff}}_0 =\begin{bmatrix}
    H^{\text{eff}}_{1}& V_1        & 0          & 0          & 0 \\
    V_1^\dagger& H_{2}      & V_2        & 0          & 0 \\
    0          & V_2^\dagger& H_{3}      & V_3        & 0 \\
    0          & 0          & V_3^\dagger& H_{4}      & V_4 \\
    0          & 0          & 0          & V_4^\dagger& H_{5}
\end{bmatrix},\]
with the blocks defined above. Finally, we expand the exponentials to leading order in the $\bm u_l$ and retain only the first harmonic in all the resulting terms of the Hamiltonian.

\begin{table}[!t]
    \centering
    \def\arraystretch{1.1}
    \setlength\tabcolsep{1ex}
    \begin{tabular}{c|c l|l}
         Variable & Value & & Source  \\\hline
         $v_F$ & $8\cdot 10^{15}$ & \AA/s & \cite{moon_electronic_2014}\\
         $t$ & 2.78 & eV & \cite{san-jose_spontaneous_2014} \\
         $\beta$ & 2 & & \cite{san-jose_spontaneous_2014}\\
         $\epsilon$ & 5.5 (5-7) & &  \cite{kwan_moire_2025}\\\hline
         $u_0$ & 152.0 & meV & \cite{moon_electronic_2014}\\
         $v_0$ & 113.3 & meV & \cite{moon_electronic_2014}\\
         $V_N$ & -1.4& eV& \cite{san-jose_electronic_2014} \\
         $V_B$ & 3.34& eV & \cite{san-jose_electronic_2014}
    \end{tabular}
    \caption{Electronic parameters of the model. We generally use $\epsilon = 5.5$, but also consider the wider range $5-7$.}
    \label{tab:sp-model}
\end{table}

\subsection{Details of the Hartree-Fock Calculation}
We consider a layer-dependent Coulomb interaction of the form
\[V_{\text{int}} = \frac{1}{2A}\sum_{\substack{\bm q,\, l,l'}} V_{ll'}(\bm q)\; :\rho_l(\bm q)\rho_{l'}(-\bm q):
,\]
where 
\[\rho_l(\bm q)=\sum_{\substack{\bm k,\,\alpha,\xi}}c^{\dagger}_{\alpha,l,\xi}(\bm k+\bm q)c_{\alpha,l,\xi}(\bm k)\]
is the Fourier component of the layer-resolved density operator in second quantization (momenta are not restricted and $:\,:$ indicates that operators are normal ordered in the previous equation), and
\[V_{ll'}(\bm q) = \frac{e^2}{2\epsilon_0 \epsilon q} \left[\frac{e^{-q(z_l z_{l'})} (-e^{2q(d_{sc}+z_l+z_{l'}}-e^{2qd_{sc}}+e^{2qz_l}+e^{2qz_{l'}})}{e^{4q d_{sc}}-1} - e^{-q|z_l - z_{l'}|}\right]\]
is the appropriate Coulomb interaction in momentum space; we work with $d_{sc} = 100$ \AA{} \cite{kwan_moire_2025}. 

To perform a Hartree-Fock decoupling of this interaction term, we find it convenient to introduce layer-resolved momentum-boost matrix operators $\hat{O}_l(\bm q)$, with matrix elements (repeated indices are not summed)
\[\left[\hat{O}_l(\bm q)\right]_{(\bm k_1,\alpha_1,l_1,\xi_1),(\bm k_2,\alpha_2,l_2,\xi_2)}=\delta_{\alpha_1,\alpha_2}\delta_{l_1,l}\delta_{l_2,l}\delta_{\xi_1,\xi_2}\delta_{\bm k_1,\bm k_2+\bm q},\]
in terms of which the density operator can be written as
\[\rho_l(\bm q)=\sum_{\substack{\bm k_1,\,\alpha_1,l_1,\xi_1}}\sum_{\substack{\bm k_2,\,\alpha_2,l_2,\xi_2}}\left[\hat{O}_l(\bm q)\right]_{(\bm k_1,\alpha_1,l_1,\xi_1),(\bm k_2,\alpha_2,l_2,\xi_2)}c^{\dagger}_{\alpha_1,l_1,\xi_1}(\bm k_1)c_{\alpha_2,l_2,\xi_2}(\bm k_2).\]
The Hartree-Fock decoupling amounts to the substitution $H_0^{\textrm{eff}}+V_{\textrm{int}}\rightarrow H^{HF}-\langle \Sigma^H\rangle/2-\langle \Sigma^F\rangle/2$, where the Hartree-Fock Hamiltonian is $H^{HF}=H_0^{\textrm{eff}}+\Sigma^H+\Sigma^F$, with Hartree and Fock potentials given in the same matrix notation as
\[\hat{\Sigma}^H = \frac{1}{A}\sum_{\substack{\bm q,\,l,l'}} V_{ll'}(\bm q)\, \hat{O}_{l}(\bm q)\,\mathrm{Tr}[\hat{N}\cdot\hat{O}_{l'}(-\bm q)]  \qquad \text{and} \qquad \hat{\Sigma}^F = -\frac{1}{A}\sum_{\substack{\bm q,\, l,l'}} V_{ll'}(\bm q)\, \hat{O}_l(\bm{q}) \cdot\hat{N}\cdot \hat{O}_{l'}(-\bm q),\]
respectively, where $\hat{N}$ are correlation matrices whose elements correspond to
\[\left[\hat{N}\right]_{(\bm k_1,\alpha_1,l_1,\xi_1),(\bm k_2,\alpha_2,l_2,\xi_2)}=\left\langle c^{\dagger}_{\alpha_2,l_2,\xi_2}(\bm k_2)c_{\alpha_1,l_1,\xi_1}(\bm k_1) \right\rangle.\]

Self-consistency relations follow from computing these expectations values with the Hartree-Fock Hamiltonian. In our calculations, we only consider correlation matrices commensurate with the moir\'e period, i.e., the only matrix elements of $\hat{N}$ different from $0$ are those with $\bm k_1$ and $\bm k_2$ separated by a moir\'e reciprocal vector $\bm G$. Starting from the single-electron correlation matrix, we introduce valley polarization and iterate these equations until convergence. The resulting single-electron spectrum is folded onto the moir\'e Brillouin zone. We employ a $\bm k$-mesh of $18\times18$ points within the moir\'e Brillouin zone, extended to the first three stars of $\bm G$ vectors.

Also, to make the calculation numerically more feasible, we introduce active band projection by assuming that only a finite set of bands near the Fermi level are significantly modified by interactions. This is captured by defining frozen and active band projection operators $\hat{P}_F$ and $\hat{P}_A = 1-\hat{P}_F$ and splitting the full density matrix $\hat{N}$ as $\hat{N}_A = \hat{P}_A\cdot \hat{N}\cdot \hat{P}_A$ and $\hat{N}_F = \hat{N} - \hat{N}_A$, where $\hat{N}_F$ is now constant (i.e. it's not updated on each iteration). We consider three bands above and three bands below the Fermi level. The calculations are performed at a fixed number of electrons per moir\'e cell.

Finally, to remove possible double-counting of interaction effects and properly set the system's vacuum, we employ the standard charge neutrality subtraction scheme in which the interaction self-energies from the charge-neutral system are removed. This can effectively be implemented by defining a new frozen density matrix
\[\hat{N}_F^{\textrm{CN}}=\hat{N}_F-\hat{N}_{\textrm{CN}}\left(=-\hat{P}_A\cdot\hat{N}_{CN}\cdot\hat{P}_A\right).\]
Note that this operator has only support on the subspace of active bands. The expressions for the projected Hartree and Fock self-energies read, respectively, as
\[\hat{\Sigma}^{H, \text{CN}}_A = \frac{1}{A}\sum_{\substack{\bm q,\, l,l'}} V_{ll'}(\bm q)\,\hat{P}_A \cdot \hat{O}_{l}(\bm q) \cdot\hat{P}_A\, \mathrm{Tr}[(\hat{N}_A + \hat{N}_F^{\text{CN}})\cdot \hat{O}_{l'}(-\bm q)]\]
and
\[\quad \Sigma^{F, \text{CN}}_A = -\frac{1}{A}\sum_{\bm q,\,l,l'} V_{ll'}(\bm q) \,\hat{P}_A\cdot\hat{O}_l(\bm q)\cdot (\hat{N}_A + \hat{N}_F^{\text{CN}})\cdot \hat{O}_{l'}(-\bm q) \cdot \hat{P}_A.\]

\subsection{Calculation of Chern Numbers}
We use two independent methods to calculate the Chern numbers presented in this article. One is direct integration of the Berry curvature over the Brillouin zone to give a approximately quantized Chern number. However, such direct integration often requires a very dense $\bm k$-mesh, and can give ambiguous answers.


The other is to exploit the $C_3$ symmetry of the mode to directly calculate the exact Chern number modulo 3, following the methods of Ref.~\cite{PhysRevB.86.115112}. At the $\gamma$-point in valley $\xi$ we have the layer-dependent representation
\[[C_3^\xi]_{l} = e^{-\xi \pi i \sigma_z/3} e^{2\pi i l/3}. \]
Note that the representation is layer-dependent because the 3-fold rotation axis is common to all layers, but there are relative shifts between them as represented in Ref.~\ref{fig:geometry}. At the $\kappa$ and $\kappa'$ points, an additional momentum boost must be included to map the rotated momentum back to the original point. Due to the numerical momentum cutoff, this results in the $C_3$ symmetry not being exact at these points, however, we numerically find the error to be negligible.

\subsection{Electrostatic Effects}


\begin{figure}[t]
    \centering
    \includegraphics[width=0.6\linewidth]{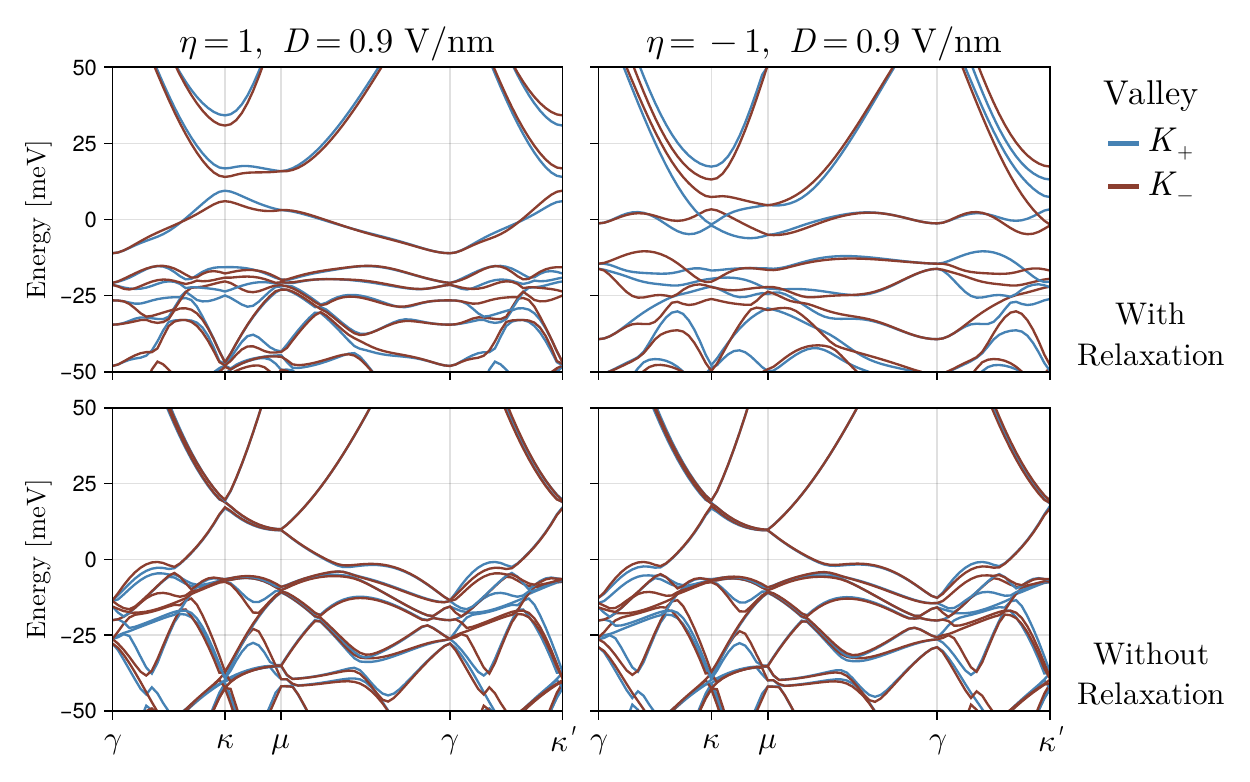}
    \caption{Relaxation helps isolate the filled conduction band in the Hartree band structure at $D = 0.9\; \text{V}/\text{nm}$ and filling $\nu=1$. This isolation is greater for $\eta = 1$, but the band is slightly flatter for $\eta=-1$. Without relaxation, both stacking configurations are similar, with the filled conduction band not being isolated. A $24\times 24$ $k$-mesh and $\epsilon = 6$ is used.}
    \label{fig:hartree_bands}
\end{figure}

In the moir\'e distant regime, while the conduction band electrons which experimentally host fractional states are located far from the moir\'e contact layer, the filled valence states are pushed towards the contact layer. Thus, one would expect the precise effects of the moir\'e to still play an important role via the long-range electron interactions. Thus, to further examine the relevance of relaxation and the detailed effects of the moir\'e potential, we initially consider electrostatic effects by including an interaction decoupling only in the density (Hartree) channel. Such interactions should partially screen the strong displacement fields, reducing the band gap but also potentially helping to isolate the first conduction band. Matching the relevant experimental conditions, we focus on the case of filling $\nu=1$ electron per moir\'e unit cell and displacement field $D = 0.9$ V/nm, with our results shown in Fig.~\ref{fig:hartree_bands}. In contrast to the full Hartree-Fock calculation above, in these calculations we do not need to choose subsets of active and frozen bands, allowing all bands to renormalize under the Hartree interaction. 

In contrast to the single-particle bands, we find that the unrelaxed system is not gapped when electrostatic interactions are included---with the first conduction band touching both the valence and higher conduction bands---indicating strong screening of the displacement field. Moreover, though there is some local valley splitting in the valence bands, the conduction band of interest remains nearly identical for both valleys. 

With relaxation included, however, the first conduction band becomes more isolated in both stacking configurations---with it being fully gapped for $\eta=1$ and only having an isolated crossing for $\eta = -1$. The strong valley splitting also persists, especially in the $\eta =-1$ case. Extending beyond the single-particle level, even the first conduction band, which previously did not display significant valley dependence, is valley split at the Hartree level. 
The conduction band in both stacking configurations fails to explicitly display a topological character in this regime, carrying $C=0$. Regardless, the fact that relaxation effects are needed to isolate the band when electrostatic interactions are considered is still a promising indication of its importance.

\begin{figure}[t]
    \centering
    \includegraphics[width=0.55\linewidth]{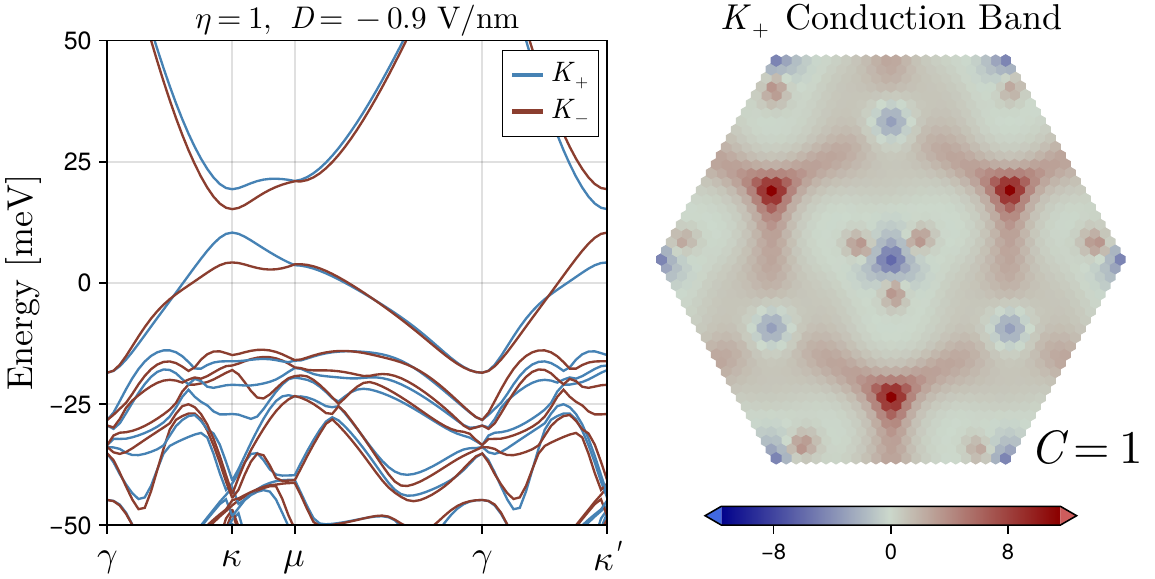}
    \caption{In the moir\'e proximal regime, relaxation effects create a topological band, with $C=1$, even at large displacement fields and with pure Hartree effects. Due to electrostatic screening, the bands are qualitatively similar between the $D>0$ and $D<0$ cases. The Berry curvature color scale is given by $\Omega({\bf k}) \cdot A_{\text{MBZ}}/(2\pi)$}
    \label{fig:hartree_chern}
\end{figure}

In the moir\'e proximal regime with $D<0$, we find that a combination of relaxation and electrostatic effects does induce a topological filled conduction band with $C=1$, as shown in Fig.~\ref{fig:hartree_chern}. As with the moir\'e distant regime, the unrelaxed bandstructure is not gapped, indicating that relaxation can induce topological character even when accounting for electrostatic renormalization in very strong displacement fields. While not the experimentally most relevant regime, these results are again indicative of the importance of properly including relaxation effects. 

\subsection{Changes with Dielectric Constant}
We use a value of $\epsilon = 5.5$ through the main text, matching previous works \cite{kwan_moire_2025}. Varying $\epsilon$---which corresponds to varying the strength of interactions---does result in quantitatively different results. In Fig.~\ref{fig:vareps} below, we compare the cases of $\epsilon = 5$ and $\epsilon = 7$, corresponding to strong and weak interactions respectively. In the former case, we find the same Chern bands with relaxation, but additionally find the unrelaxed first conduction band to be fully occupied.  In contrast, in the $\epsilon = 7$ case, we find the first conduction band to be unfilled in both the relaxed and unrelaxed scenarios. Moreover, the $\eta = 1$ relaxed band has developed a Chern number $C=2$, with a band crossing occurring at the $\kappa$ point around $\epsilon\approx6$. These results show that strong interactions are needed to develop flat bands with the experimentally observed Chern number, with relaxation playing an important auxiliary role in flattening the bands. 

\begin{figure}[!b]
    \centering
    \subfloat{
        \includegraphics[width=0.48\textwidth]{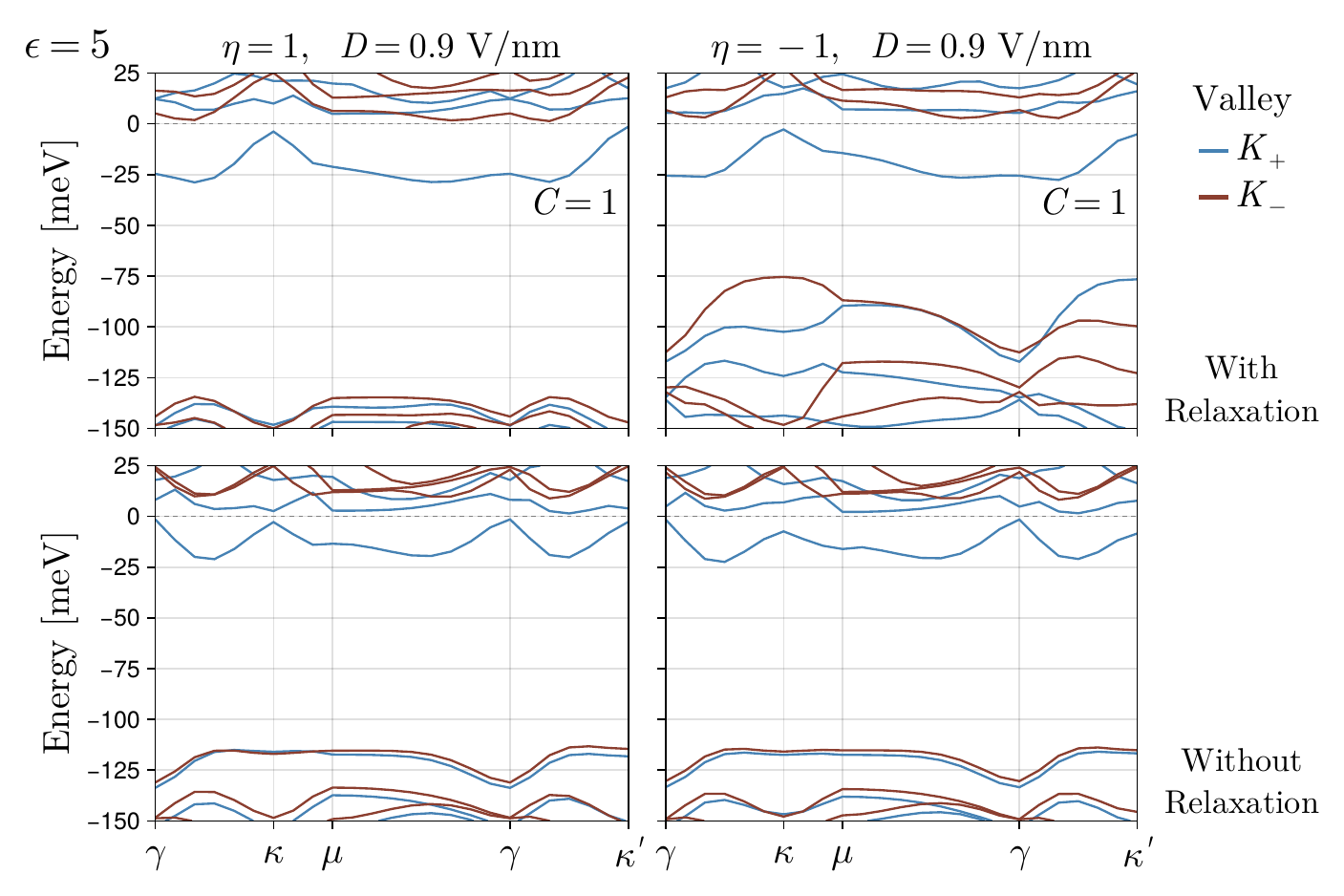}
    }
    \hfill
    \subfloat{
        \includegraphics[width=0.48\textwidth]{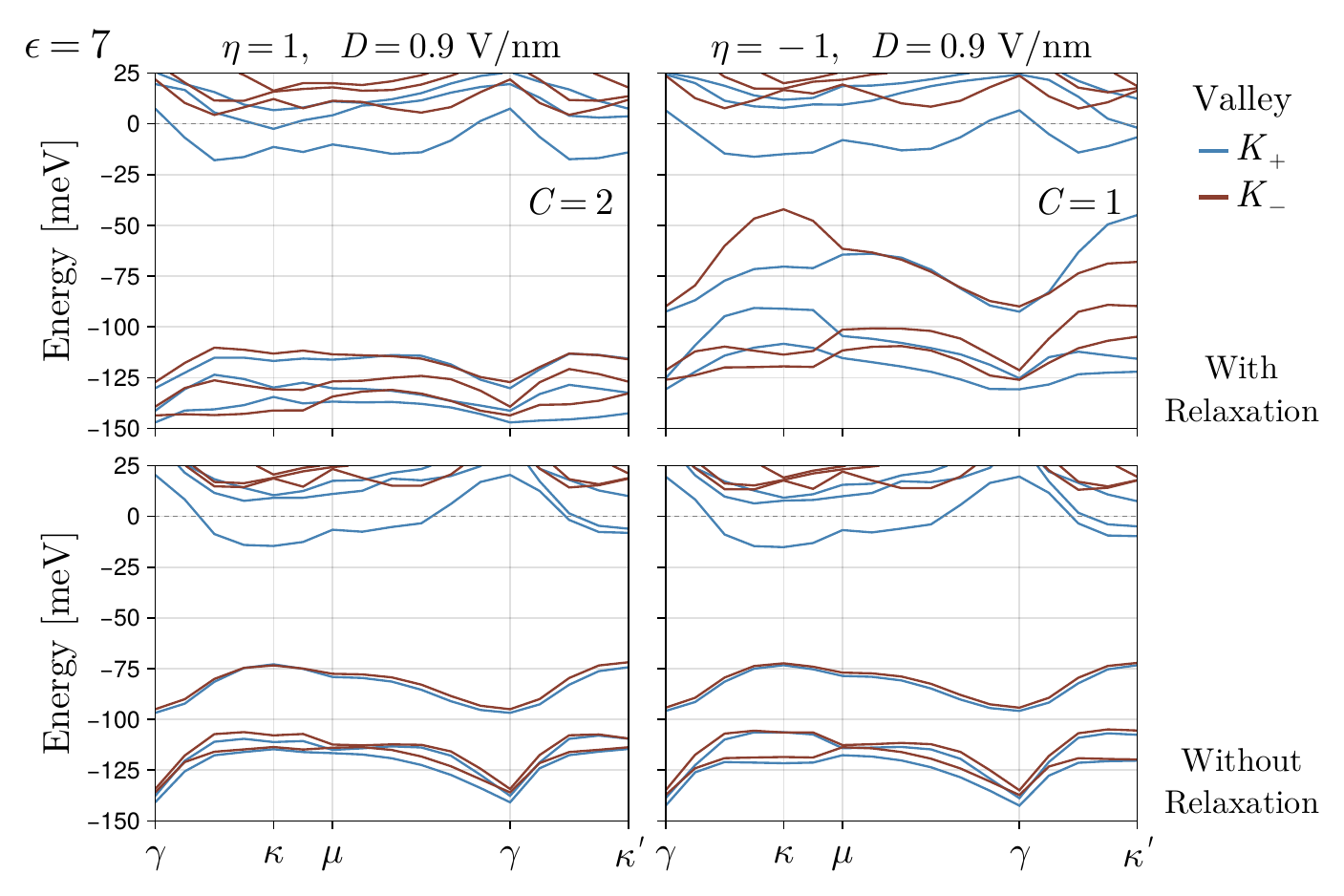}
    }
    \caption{The $\epsilon = 5$ (left) and $\epsilon = 7$ (right) Hartree-Fock band structures. An $18\times 18$ $k$-mesh is used for $\epsilon = 5$, while a $12\times 12$ $k$-mesh is used for $\epsilon = 7$; all other parameters are the same as in the main text. }
    \label{fig:vareps}
\end{figure}

\end{document}